\documentclass[12pt]{article}
\usepackage{amssymb,amsmath,epsfig}

\begin{document}

\title{\bf Impact of Extended Starobinsky Model on Evolution of
Anisotropic, Vorticity-free Axially Symmetric Sources}

\author{Ifra Noureen \thanks{ifra.noureen@gmail.com} ${}^{(a)}$, Abdul Aziz Bhatti
\thanks{drabhatti@umt.edu.pk} ${}^{(b)}$,  M. Zubair
\thanks{drmzubair@ciitlahore.edu.pk} ${}^{(c)}$, \\
${}^{(a,b)}$ School of Science and Technology,\\University of
Management and Technology, Lahore-54000, Pakistan.\\ ${}^{(c)}$
Department of Mathematics,\\ COMSATS Institute of Information
Technology, Lahore-54700, Pakistan.}
\date{}
\maketitle

\begin{abstract}
We study the implications of $R^n$ extension of Starobinsky model on
dynamical instability of Vorticity-free axially symmetric
gravitating body. The matter distribution is considered to be
anisotropic for which modified field equations are formed in context
of $f(R)$ gravity. In order to achieve the collapse equation, we
make use of the dynamical equations, extracted from linearly
perturbed contracted Bianchi identities. The collapse equation
carries adiabatic index $\Gamma$ in terms of usual and dark source
components, defining the range of stability/insatbility in Newtonian
(N) and post-Newtonian (pN) eras. It is found that supersymmetric
supergravity $f(R)$ model represents the more practical substitute
of higher order curvature corrections.
\end{abstract}


\section{Introduction}

The compact gravitating objects are commonly studied by assuming
spherical symmetry. The observational signatures suggests that self
gravitating compact sources may deviate from spherical symmetry.
These incident deviations from spherical symmetry provides scope for
the study of axially symmetric gravitating systems. However, such
line element in accordance with Weyl gauge largely constrains the
class of possible sources (static/dynamic) \cite{1}. Although the
consideration of non-static sources and angular momentum leave
complicated analysis, but lack of the spherical symmetry in most
realistic scenario must be considered. Thus, the explorations
regarding outcome of stellar evolution with assumption of axial
symmetry is worthwhile.

The analysis of stability range plays significant role in relevant
fields, such as astronomy, astrophysics and structure formation
theories. The gravitating bodies remain stable as long as the
balance is being maintained between outward drawn pressure induced
by internal fusion and the inward gravitational pull \cite{2}. The
stellar collapse originates from the situations where gravity
dominates as a consequence of internal fuel consumption. Instability
range of compact sources varies along with their mass, supermassive
stars has tendency of wider instability range implying smaller life
span \cite{3}, while stars having mass of the order of one solar
mass tends to be more stable.

The initial contribution on dynamical instability was of
Chandrasekhar \cite{4}, he described a pattern for the
representation of instability range in terms of adiabatic index for
ideal matter distribution. Hillebrandt and Steinmetz \cite{5} found
the instability criterion for compact objects having pressure
anisotropies in matter configuration. Many authors
\cite{6}-\cite{16}, emphasized on stellar evolution and stability
range for a number of matter distributions, such as isotopic fluid,
anisotropy, zero expansion, shear-free condition, radiation and
dissipation. The results obtained from their analysis established a
major remark that nominal variations in fluid configuration alters
stability range significantly. Also, the high radiation transport
leaves instabilities in the system.

Modified theories have gained more attention to count with the issue
of cosmic acceleration \cite{16a}. Likewise in general relativity
(GR), the instability problem has also been widely discussed in
modified gravity theories namely, $f(R)$, $f(R,T)$, where $T$ is the
trace of energy momentum tensor, $f(G)$, Brans-Dicke theory etc. The
modified theories provide higher order corrections to GR on large
scale structures for inclusion of dark energy substitutes. People
\cite{17}-\cite{23}, worked out the instability problem in $f(R)$
theory for various matter distributions with and without Maxwell
source, concluding that the inclusion of higher order curvature
terms depicts the more broader picture of factors affecting the
stability. The evolution of compact bodies in Gauss-Bonnet ($f(G)$)
and $f(\mathcal{T})$ (where $\mathcal{T}$ is the torsion scalar)
theory has been explored in \cite{24}-\cite{26}. Recently, the
dynamics of isotropic and anisotropic fluid has been studied in
$f(R, T)$ theory \cite{27}-\cite{28}.

$f(R)$ gravity represents the elementary modification in GR,
including higher order curvature terms in Einstein-Hilbert (EH)
action. In $f(R)$, the Ricci scalar $R$ modifies to $\sqrt{-g}f(R)$
in EH action incorporating higher curvature terms
\cite{29}-\cite{32}, when $g$ is metric tensor and $f(R)$ stands for
the general function of $R$. The EH action in $f(R)$ is of the form
\begin{equation}\label{1a}
S_{f(R)}=\frac{1}{2\kappa}\int d^{4}x\sqrt{-g}f(R)+S,
\end{equation}
where $\kappa$ stands for the coupling constant and $S$ is the
action for usual matter. Any specific form of $f(R)$ is viable, if
it is in accordance with the viability criterion, i.e., the second
order derivative of considered model must be positive along with
realistic situations such as weak lensing, cosmic microwave
background and clustering spectrum \cite{33}-\cite{36}.

In GR, the anisotropic dissipative and shear-free fluid models have
been studied for the dynamic axially and reflection symmetric
sources \cite{37, 38}, ignoring rotation ($dtd\phi$ term) in general
axial symmetry. Sharif and Zousaf \cite{39} explained the dynamics
of restricted non-static axially symmetric collapse for Starobinsky
model \cite{39a} in anisotropic environment. Herein, we are aiming
to discuss the impact of supersymmetric Starobinsky model
constructed in \cite{40} on dynamics of the axially symmetric
gravitating source. The components of modified field equations are
used to construct the conservation equations for anisotropic matter
configuration. We have implemented the perturbation approach in
order to develop the collapse equation and discuss the role of usual
matter and dark source components in establishment of
stabilit/instability range. Perturbation of dynamical equations lead
to the collapse equation which is further used to discuss
instability range in terms of $\Gamma$ for N and pN regimes.

The manuscript arrangement is: Section \textbf{2} contains the
components of field equations and conservation equations for axially
symmetric self gravitating objects along with the description of
anisotropic matter configuration. The $f(R)$ model is furnished in
section \textbf{3} along with the perturbed field equations and
Bianchi identities. The collapse equation and dynamical analysis in
N and pN eras is provided in section \textbf{4}. The last section
constitutes the conclusion followed by an appendix.

\section{Dynamical Equations}

For the dynamical analysis, we consider the spacetime which
describes the restricted non-static axial symmetry avoiding the
terms of reflection and rotation about the the symmetry axis. The
reduced form of general axially symmetric spacetime in spherical
coordinates is \cite{41}
\begin{equation}\label{1}
ds^2=-A^2(t,r,\theta)dt^{2}+B^2(t,r,\theta)dr^{2}+B^2(t,r,\theta)r^2
d\theta^{2} +C^2(t,r,\theta)d\phi^{2}.
\end{equation}
It is worthwhile to mention here that, this analysis belongs to the
restricted class of axially symmetric sources, i.e., absence of
meridional motions and vorticity. In general axial symmetry, five
independent metric functions should appear in the line element, due
to restricted character (excluding meridional motions and motions
around symmetry axis), we have three of them. we are dealing with
analytic approach to present the dynamical analysis of gravitating
source containing three independent metric functions. Already, it is
a cumbersome task to analyze the system without reflection and
rotation then rather to include such terms so we have neglected the
terms $dtd\theta$ and $dtd\phi$ in the general axially symmetric
line element.

The gravitating source is considered to have anisotropic matter
configuration defined by the energy momentum tensor \cite{1}
\begin{eqnarray}\nonumber&&
T_{uv}=(\rho+p_\perp)V_{u}V_{v}-(K_uK_v-\frac{1}{3}h_{uv})(P_{zz}-P_{xx})-
(L_uL_v-\frac{1}{3}h_{uv})(P_{zz}\\\label{3}&&-P_{xx})+Pg_{uv}+2K_{(u}L_{v)}P_{xy},
\end{eqnarray}
where
\begin{equation}\nonumber
P=\frac{1}{3}(P_{xx}+P_{yy}+P_{zz}), \quad h_{uv}=g_{uv}+V_{u}V_{v},
\end{equation}
$P_{xx}, P_{yy}, P_{zz}$ and $P_{xy}$ denote different stresses
inducing pressure anisotropy, provided that $P_{xy}=P_{yx}$ and
$P_{xx}\neq P_{yy}\neq P_{zz}$. The energy density is labeled as
$\rho$, $V_{u}$ is for four-velocity, $K_u$  and $L_u$ denote four
vectors in radial and axial directions respectively. We have chosen
Eulerian frame to describe the quantities, implying that
\begin{equation}\label{4}
V_{u}=-A\delta^{0}_{u},\quad K_{u}=B\delta^{1}_{u}1,\quad
L_{u}=rB\delta^{2}_{u}.
\end{equation}

The variation of EH action (\ref{1a}) with respect to metric tensor
$g_{uv}$ yields the following field equations \cite{23}
\begin{equation}\label{5}
f_RR_{uv}-\frac{1}{2}f(R)g_{uv}-\nabla_{u}
\nabla_{v}f_R+ g_{uv} \Box f_R=\kappa T_{uv},\quad(u, v =0,1,2,3),
\end{equation}
where $\nabla_{u}$ is covariant derivative,
$\Box=\nabla^{u}\nabla_{v}$, $f_R\equiv df(R)/dR$. Likewise GR, we
may write field equations as
\begin{equation}\label{6}
G_{uv}=\frac{\kappa}{f_R}[\overset{(D)}{T_{uv}}+T_{uv}]=
\frac{\kappa}{f_R}\overset{(D)}{T_{uv}}+\frac{\kappa}{f_R}T_{uv},
\end{equation}
where $\overset{(D)}{T_{uv}}$ represents energy momentum tensor of
dark source contribution given by
\begin{equation}\label{7}
\overset{(D)}{T_{uv}}=\frac{1}{\kappa}\left[\frac{f(R)-Rf_R}{2}g_{uv}+\nabla_{u}
\nabla_{v}f_R -g_{uv} \Box f_R\right].
\end{equation}
The non-zero components of modified (effective) Einstein tensor for
axial symmetry takes the following form
\begin{eqnarray}\nonumber
G^{00}&=&\frac{\kappa}{A^2f_R}\rho+\frac{1}{A^2f_R}\left[\frac{f-Rf_R}{2}
-\frac{\dot f_R}{A^2}\left(\frac{\dot{2B}}{B}+\frac{\dot{C}}{C}\right)
-\frac{f_R'}{B^2}\left(\frac{2B'}{B}-\frac{C'}{C}+\frac{1}{r}\right)\right.\\\label{8}&&\left.
-\frac{f_R^\theta}{r^2B^2}\left(\frac{2B^\theta}{B}-\frac{C^\theta}{C}\right)
+\frac{f_R''}{B^2}\right],
\\\label{9}
G^{01}&=&\frac{1}{A^2B^2f_R}\left[\dot{f_R}'
-\frac{A'}{A}\dot{f_R}-\frac{\dot{B}}{B}f_R'\right],
\\\label{10}
G^{02}&=&\frac{1}{r^2A^2B^2f_R}\left[\dot{f_R}^\theta
-\frac{A^\theta}{A}\dot{f_R}-\frac{\dot{B}}{B}f_R^\theta\right],
\end{eqnarray}
\begin{eqnarray}\nonumber
G^{11}&=&\frac{\kappa}{B^2f_R}P_{xx}+\frac{1}{B^2f_R}\left[-
\frac{f-Rf_R}{2}- \frac{\ddot{f_R}}{A^2}-\frac{
f_R^{\theta\theta}}{r^2B^2}-\frac{\dot f_R}{A^2}
\left(\frac{\dot{A}}{A}-\frac{\dot{B}}{B}+\frac{\dot{C}}{C}\right)
\right.\\\label{11}&&\left.-\frac{f_R'}{B^2}\left(\frac{A'}{A}-\frac{B'}{B}
+\frac{C'}{C}-\frac{1}{r}\right)-
\frac{f_R^\theta}{r^2B^2}\left(\frac{A^\theta}{A}-\frac{3B^\theta}{B}
+\frac{C^\theta}{C}\right)\right],
\\\label{12}
G^{12}&=&\frac{\kappa}{r^2B^4f_R}P_{xy}+\frac{1}{r^2B^4f_R}\left[f_R'^\theta-
\frac{B^\theta}{B}f'_R-\frac{B'}{B}f_R^\theta\right],
\\\nonumber
G^{22}&=&\frac{\kappa}{r^2B^2f_R}P_{yy}+\frac{1}{r^2B^4f_R}\left[-\frac{f-Rf_R}{2}+
\frac{\ddot{f_R}}{A^2}-\frac{\dot f_R}{A^2}\left(\frac{\dot{A}}{A}
-\frac{\dot{B}}{B}-\frac{\dot{C}}{C}\right)\right.\\\label{13}&&\left.-\frac{f_R''}{B^2}
-\frac{f_R'}{B^2}\left(\frac{A'}{A}-\frac{B'}{B}+\frac{C'}{C}\right)
-\frac{f_R^\theta}{r^2B^2}\left(\frac{A^\theta}{A}-\frac{B^\theta}{B}
+\frac{C^\theta}{C}\right)\right], \\\nonumber
G^{33}&=&\frac{\kappa}{C^2f_R}P_{zz}+\frac{1}{C^2f_R}\left[-\frac{f-Rf_R}{2}+
\frac{\ddot{f_R}}{A^2}-\frac{ f_R^{\theta\theta}}{r^2B^2}-\frac{\dot
f_R}{A^2} \left(\frac{\dot{A}}{A}
-\frac{\dot{2B}}{B}\right)\right.\\\label{14}&&\left.
-\frac{f_R''}{B^2}-\frac{f_R'}{B^2}\left(\frac{A'}{A}-\frac{2B'}{B}-\frac{1}{r}\right)
-\frac{f_R^\theta}{r^2B^2}\left(\frac{A^\theta}{A}-\frac{2B^\theta}{B}\right)\right].
\end{eqnarray}
Herein dot, prime and $\theta$ indicates the time, radial and axial
derivatives respectively. The Ricci scalar corresponding to the
metric is
\begin{eqnarray}\nonumber
R&=&\frac{2}{A^2}\left[\frac{\dot{A}}{A}\left(\frac{\dot{2B}}{B}+\frac{\dot{C}}{C}\right)-
\frac{\dot{B}}{B}\left(\frac{\dot{B}}{B}+\frac{\dot{2C}}{C}\right)-\frac{2\ddot{B}}{B}
-\frac{\ddot{C}}{C}\right]\\\nonumber&&+\frac{2}{B^2}\left[\frac{A''}{A}+
\frac{A'C'}{AC}+\frac{B''}{B}-\frac{1}{r}\left(\frac{A'}{A}-\frac{B'}{B}-\frac{C'}{C}\right)-
\frac{B'^2}{B^2}\right.\\\label{14'}&&\left.+\frac{C''}{C}+\frac{1}{r^2}
\left(\frac{A^{\theta\theta}}{A}+
\frac{B^{\theta\theta}}{B}+\frac{C^{\theta\theta}}{C}
-\frac{{B^\theta}^2}{B^2}+\frac{A^\theta
C^\theta}{AC}\right)\right].
\end{eqnarray}

The dynamical equations play pivotal role in description of
evolution, conservation provides basis for the collapse equation. In
order to formulate collapse equation, we first need to develop
dynamical equations by taking contracted Bianchi identities as under
\begin{eqnarray}\label{15}
&&G^{uv}_{;v}V_{u}=0 \Rightarrow \left[\frac{\kappa}{f_R}T^{0v}
+\frac{\kappa}{f_R}\overset{(D)}{T^{0v}}\right]_{;v}(-A)=0,
\\\label{16}
&&G^{uv}_{;v}K_{u}=0 \Rightarrow \left[\frac{\kappa}{f_R}T^{1v}
+\frac{\kappa}{f_R}\overset{(D)}{T^{1v}}\right]_{;v}(B)=0,
\\\label{17}
&&G^{uv}_{;v}L_{u}=0 \Rightarrow \left[\frac{\kappa}{f_R}T^{2v}
+\frac{\kappa}{f_R}\overset{(D)}{T^{2v}}\right]_{;v}(rB)=0,
\end{eqnarray}
implying
\begin{eqnarray}\nonumber &&
G^{00}_{,0}+ G^{01}_{,1}+G^{02}_{,2}+G^{00}\left(\frac{2\dot{A}}{A}
+\frac{\dot{2B}}{B}+\frac{\dot{C}}{C}\right)+G^{01}\left(\frac{3A'}{A}+\frac{2B'}{B}
+\frac{C'}{C}+\frac{1}{r}\right)\\\label{18}&&+G^{02}\left(\frac{3A^\theta}{A}+\frac{2B^\theta}{B}
+\frac{C^\theta}{C}\right)+G^{11}\frac{B\dot{B}}{A^2}+G^{22}\frac{r^2B\dot{B}}{A^2}
+G^{33}\frac{C\dot{C}}{A^2}=0,
\end{eqnarray}
\begin{eqnarray}\nonumber &&
G^{01}_{,0}+
G^{11}_{,1}+G^{12}_{,2}+G^{00}\frac{AA'}{B^2}+G^{01}\left(\frac{\dot{A}}{A}
+\frac{\dot{4B}}{B}+\frac{\dot{C}}{C}\right)+G^{11}\left(\frac{A'}{A}+\frac{3B'}{B}
+\frac{C'}{C}\right.\\\label{19}&&\left.+\frac{1}{r}\right)+G^{12}\left(\frac{A^\theta}{A}
+\frac{4B^\theta}{B}
+\frac{C^\theta}{C}\right)-G^{22}\left(r+\frac{r^2B'}{B}\right)+G^{33}\frac{CC'}{B^2}=0,
\\\nonumber &&
G^{02}_{,0}+ G^{12}_{,1}+G^{22}_{,2}+G^{00}\frac{AA^\theta}{r^2B^2}+G^{02}\left(\frac{\dot{A}}{A}
+\frac{\dot{4B}}{B}+\frac{\dot{C}}{C}\right)-\frac{B^\theta}{r^2B}G^{11}+\left(\frac{A'}{A}
\right.\\\label{20}&&\left.+\frac{4B'}{B}
+\frac{C'}{C}+\frac{3}{r}\right)G^{12}+G^{22}\left(\frac{A^\theta}{A}+\frac{3B^\theta}{B}
+\frac{C^\theta}{C}\right)-G^{33}\frac{CC^\theta}{r^2B^2}=0.
\end{eqnarray}
The notation $0, 1$ and $2$ denotes $t, r$ and $\theta$,
respectively. The components of field equations given in Eqs.
(\ref{8})-(\ref{14}) can be inserted in above three equations to
view matter and effective components. In following section we
present the extended Starobinsky model and also the perturbation
scheme for dynamical equations.

\section{Extended Starobinsky model and Perturbation approach}

We consider the supersymmetric supergravity $f(R)$ model
representing $R^n$ extension of well known Starobinsky model
\cite{40}
\begin{equation}\label{21}
f(R)=R+\alpha R^2+\beta R^n,
\end{equation}
where $n\geq3$, $\alpha$ and $\beta$ are positive quantities,
considered to be positive for stable stellar configuration. The
perturbation scheme have been implemented to evaluate role of the
factors contributing in the establishment of instability range.
Initially all the quantities are taken to be in hydrostatic
equilibrium and time transition implicates the time dependence as
well. First order perturbations are introduced in dynamical and
field equations, assuming $0<\epsilon\ll1$
\begin{eqnarray}\label{22}
A(t,r,\theta)&=&A_0(r,\theta)+\epsilon D(t)a(r,\theta),\\\label{23}
B(t,r,\theta)&=&B_0(r,\theta)+\epsilon D(t)b(r,\theta),\\\label{24}
C(t,r,\theta)&=&C_0(r,\theta)+\epsilon D(t)c(r,\theta),\\\label{25}
\rho(t,r,\theta)&=&\rho_0(r,\theta)+\epsilon
{\bar{\rho}(t,r,\theta)},
\end{eqnarray}
\begin{eqnarray}
\label{26} P_{xx}(t,r,\theta)&=&P_{xx0}(r,\theta)+\epsilon
{\bar{P}_{xx}(t,r,\theta)},
\\\label{27}
P_{yy}(t,r,\theta)&=&P_{yy0}(r,\theta)+\epsilon {\bar{P}_{yy}(t,r,\theta)},
\\\label{28}
P_{zz}(t,r,\theta)&=&P_{zz0}(r,\theta)+\epsilon {\bar{P}_{zz}(t,r,\theta)},
\\\label{29}
P_{xy}(t,r,\theta)&=&P_{xy0}(r,\theta)+\epsilon {\bar{P}_{xy}(t,r,\theta)},
\\\label{30}
R(t,r,\theta)&=&R_0(r,\theta)+\epsilon D(t)e(r,\theta),
\\\nonumber f(R)&=&\left(R_0+\alpha
R_0^2+\beta R_0^{n}\right)+\epsilon D(t) e(r,\theta)\left(1\right.\\\label{31}
&&\left.+2\alpha
R_0+\beta nR_0^{n-1}\right),\\\nonumber f_R(R)&=&\left(1+2\alpha
R_0+\beta nR_0^{n-1}\right)+\epsilon D(t) e(r,\theta)\left(2\alpha
\right.\\\label{32}
&+&\left.\beta n(n-1)R_0^{n-2}\right).
\end{eqnarray}
The perturbed form of dynamical equations (\ref{18})-(\ref{20})
leads to the following set of equations
\begin{eqnarray}\nonumber&&
\kappa\left[\dot{\bar{\rho}} +\left\{\rho_0\left(\frac{2b}{B_0}+\frac{c}{C_0}-\frac{J}{I}\right)+
\frac{b}{B_0}(P_{xx0}+P_{yy0})+\frac{c}{C_0}P_{zz0}+\frac{Z_{1p}}{\kappa}\right\}\dot{D}\right]=0,\\\label{33}&&
\\\nonumber&&
\kappa\left[(\frac{\bar{P_{xx}}}{I})'
+(\frac{\bar{P_{xy}}}{I})^\theta+\frac{A_0'}{A_0}\bar{\rho}+D\left\{\rho_0
\left(\frac{(aA_0)'}{A_0^2}+\frac{A_0'}{A_0}\right)+P_{xx0}\left(\left(\frac{a}{A_0}\right)'
\right.\right.\right.\\\nonumber&&\left.\left.\left.
+3\left(\frac{b}{B_0}\right)'+\left(\frac{c}{C_0}\right)'\right)+
\frac{1}{r^2}P_{xy0}\left(\left(\frac{a}{A_0}\right)^\theta
+4\left(\frac{b}{B_0}\right)^\theta+\left(\frac{c}{C_0}\right)^\theta\right)
\right.\right.\\\nonumber&&\left.\left.
-{P_{yy0}}\left(\frac{b}{B_0}\right)'-P_{zz0}\left(\frac{(cC_0)'}{C_0^2}
+\frac{C_0'}{C_0}\right)\right\}
+\frac{\bar{P_{xy}}}{r^2}\left(\frac{A_0^\theta}{A_0}+\frac{3B_0^\theta}{B_0}
+\frac{C_0^\theta}{C_0}\right)
\right.\\\nonumber&&\left.+\bar{P_{xx}}\left(\frac{A_0'}{A_0}+\frac{2B_0'}{B_0}+\frac{C_0'}{C_0}
+\frac{1}{r}\right)-\bar{P_{yy}}\left(\frac{1}{r}+\frac{B_0'}{B_0}\right)
-\bar{P_{zz}}\frac{C_0'}{C_0}\right]\\\label{34}&&+\frac{Z_{2p}}{\kappa}=0,
\\
\nonumber&&
\kappa\left[\left(\frac{\bar{P_{xy}}}{I}\right)'+\left(\frac{\bar{P_{yy}}}{I}\right)^\theta
+\frac{1}{r^2}\left(\frac{A_0^\theta}{A_0}\bar{\rho}-\frac{B_0^\theta}{B_0}\bar{P_{xx}}
+\left(\frac{A_0^\theta}{A_0}+3\frac{B_0^\theta}{B_0}+\frac{C_0^\theta}{C_0}\right)\bar{P_{yy}}
\right.\right.\\\nonumber&&\left.\left.
+\frac{C_0^\theta}{C_0}\bar{P_{zz}}\right)+D\left\{\frac{1}{r^2}\left(\rho_0\left(\frac{(aA_0)^\theta}{A_0^2}
+\frac{A_0^\theta}{A_0}\right)-P_{xx0}\left(\frac{b}{B_0}\right)^\theta-P_{zz0}\left(\frac{(cC_0)^\theta}{C_0^2}
\right.\right.\right.\right. \\\nonumber&&\left.\left.\left.\left.
+\frac{C_0^\theta}{C_0}\right)+P_{yy0}\left(\left(\frac{a}{A_0}\right)^\theta
+3\left(\frac{b}{B_0}\right)^\theta
+\left(\frac{c}{C_0}\right)^\theta
\right)\right)+P_{xy0}\left(\frac{A_0'}{A_0}+\frac{4B_0'}{B_0}
\right.\right.\right.\\\label{35}&&\left.\left.\left.+\frac{C_0'}{C_0}
+\frac{3}{r}\right)\right\}\right]+\frac{Z_{3p}}{\kappa}=0,
\end{eqnarray}
where $Z_{1p}, Z_{2p}$ and $Z_{3p}$ are perturbed dark source
components of conservation equations as shown in appendix. For the
sake of simplicity, we put $I = 1+2\alpha R_0+\beta nR_0^{n-1}$, $J
= e(2\alpha R_0+\beta n(n-1)R_0^{n-2}$ and $L = \alpha R_0^2+\beta
(n-1)R_0^{n}$. Energy density $\bar{\rho}$ is derived from
Eq.(\ref{33}) as
\begin{eqnarray}\nonumber&&
\dot{\bar{\rho}}=-\left\{\rho_0\left(\frac{2b}{B_0}+\frac{c}{C_0}-\frac{J}{I}\right)+
\frac{b}{B_0}(P_{xx0}+P_{yy0})+\frac{c}{C_0}P_{zz0}+\frac{1}{\kappa}Z_{1p}\right\}D.\\\label{36}&&
\end{eqnarray}

The relationship between energy density and the corresponding
stresses can be taken as \cite{39, 43}
\begin{equation}\label{39}
\bar{P}_i=\Gamma\frac{p_{i0}}{\rho_0+p_{i0}}\bar{\rho}.
\end{equation}
Here, $\Gamma$ denotes the variation of pressure stresses depending
on energy density. Using Eqs.(\ref{36}) and (\ref{39}), we have
\begin{eqnarray}\nonumber
&&\bar{P}_{xx}=-\Gamma\frac{p_{xx0}}{\rho_0+p_{xx0}}\left\{\rho_0\left(\frac{2b}{B_0}
+\frac{c}{C_0}-\frac{J}{I}\right)+
\frac{b}{B_0}(P_{xx0}+P_{yy0})+\frac{c}{C_0}P_{zz0}\right.\\\label{40}&&\left.+\frac{1}{\kappa}Z_{1p}\right\}D,
\\\nonumber
&&\bar{P}_{yy}=-\Gamma\frac{p_{yy0}}{\rho_0+p_{xx0}}\left\{\rho_0\left(\frac{2b}{B_0}
+\frac{c}{C_0}-\frac{J}{I}\right)+
\frac{b}{B_0}(P_{xx0}+P_{yy0})+\frac{c}{C_0}P_{zz0}\right.\\\label{41}&&\left.+\frac{1}{\kappa}Z_{1p}\right\}D,
\\\nonumber
&&\bar{P}_{zz}=-\Gamma\frac{p_{zz0}}{\rho_0+p_{xx0}}\left\{\rho_0\left(\frac{2b}{B_0}
+\frac{c}{C_0}-\frac{J}{I}\right)+
\frac{b}{B_0}(P_{xx0}+P_{yy0})+\frac{c}{C_0}P_{zz0}\right.\\\label{42}&&\left.+\frac{1}{\kappa}Z_{1p}\right\}D,
\\\nonumber
&&\bar{P}_{xy}=-\Gamma\frac{p_{xy0}}{\rho_0+p_{xx0}}\left\{\rho_0\left(\frac{2b}{B_0}
+\frac{c}{C_0}-\frac{J}{I}\right)+
\frac{b}{B_0}(P_{xx0}+P_{yy0})+\frac{c}{C_0}P_{zz0}\right.\\\label{43}&&\left.+\frac{1}{\kappa}Z_{1p}\right\}D.
\end{eqnarray}

An ordinary differential equation can be extracted from perturbed
form of Ricci scalar as
\begin{equation}\label{37}
\ddot{D}(t)-Z_4(r) D(t)=0,
\end{equation}
where $Z_4$ describes the perturbed quantities and is given in
appendix, defined in a way that positivity holds for each term, the
solution of above equation becomes
\begin{equation}\label{38}
D(t)=-e^{\sqrt{Z_4}t}.
\end{equation}
The set of dynamical equations (\ref{33})-(\ref{35}) assists in
formation of collapse equation. Eq.(\ref{33}) has been used to
extract expression for perturbed energy density that is further
utilized to determine the pressure anisotropy. Insertion of
perturbed quantities in Eqs.(\ref{34}) and (\ref{35}) serve as the
evolution equation that are identical along radial and axial
coordinates and any of them can be employed to discuss stellar
evolution.

\section{Dynamical Analysis of N and pN Approximation}

Herein, we will use Eq.(\ref{34}) as collapse equation to estimate
instability range for N and pN regions by substitution of perturbed
quantities found in Eqs.(\ref{36}), (\ref{38}) and
(\ref{40})-(\ref{43}). In the following subsections we discuss the
dynamical analysis in N and pN approximation.

\subsection{Newtonian limit}

To arrive at N-approximation, we substitute $A_0=1,~B_0=1$,
$\rho_0\gg p_{i0}; i=xx, yy, xy, yy$ and without loss of generality
$C_0=r$, so Eq.(\ref{34}) implies

\begin{equation}
\Gamma <
\frac{a'\rho_{0}-P_{yy0}b'-\frac{P_{zz0}}{r^2}\left(c'+r\right)+U_1+
\frac{1}{\kappa}Z_{2^{N}_{p}}}{\frac{1}{r}\left(2b+c-\frac{J}{I}\right)
\left(2P_{xx0}+P_{yy0}+P_{zz0}\right)+U_2},
\end{equation}
where $Z_{2^{N}_{p}}$ are the terms of $Z_{2p}$ that belongs to
N-approximation, and
\begin{eqnarray}\nonumber
U_1&=&P_{xx0}(a'-3b'-\frac{1}{r^2})+P_{xy0}(a^\theta+4b^\theta),
\\\nonumber
U_2&=&\left\{\frac{P_{xx0}}{I}(2b+c-\frac{J}{I})\right\}'
+\left\{\frac{P_{xy0}}{I}(a+4b)\right\}^\theta.
\end{eqnarray}
The inequality mentioned above express $\Gamma$ in terms of the
perturbed metric and dark source terms. The gravitating system
remains unstable as long as the above condition holds. Each term
belonging to inequality is presumed in a way that whole expression
on right side of the $\Gamma$ remains positive. Since the perturbed
stresses are negative depicting the collapsing scenario, so in order
to satisfy the above inequality some physical quantities are
constrained as follows
\begin{equation}
\frac{J}{I}>2b+c, \quad a<-4b, \quad a'<3b'+\frac{1}{r^2}.
\end{equation}
These restrictions are vital in establishment of the instability
range in N-approximation.

\subsection*{Post Newtonian Regime}

In this approximation we take $A_0=1-\frac{m_0}{r}$ and
$B_0=1+\frac{m_0}{r}$ implying
\begin{equation}
\Gamma < \frac{\frac{\rho_0}{r-m_0}U_4+U_5-\frac{m_0}{r(r-m_0)\kappa}Z_{1^{N}_{p}}
+\frac{1}{\kappa}Z_{2^{N}_{p}}}{U_6
+\left(\frac{P_{xx0}}{I}U_5\right)'+\left(\frac{P_{xy0}}{I}U_5\right)^\theta},
\end{equation}
Where
\begin{eqnarray}\nonumber
U_3&=&\left(\frac{2br}{r+m_0}+\frac{c}{r}-\frac{J}{I}\right), \quad U_4=a'r
+\frac{m_0}{r-m_0}(a+1)-\frac{m_0}{r}U_3,
\\\nonumber
U_5&=&P_{xx0}\left\{\left(\frac{ar}{r-m_0}\right)'+3\left(\frac{br}{r+m_0}\right)'
+\left(\frac{c}{r}\right)'\right\}+\frac{P_{xy0}}{r^2}\left\{\left(\frac{ar}{r-m_0}\right)^\theta
\right.\\\nonumber &&\left.+4\left(\frac{br}{r+m_0}\right)^\theta\right\}-P_{yy0}
\left(\frac{br}{r+m_0}\right)'-
\frac{P_{zz0}}{r}\left(\frac{c}{r}+c'+1\right),
\\\nonumber
U_6&=&\frac{U_3}{r^2}\left\{2m_0P_{xy0}\left(\frac{2m_0-r}{r^2-m_0^2}\right)+
rP_{xx0}\left(2+\frac{m_0(3m_0-r)}{r^2-m_0^2}\right)+P_{zz0}
\right.\\\nonumber&&\left.+P_{yy0}\left(\frac{r-2m_0}{r-m_0}\right)\right\}.
\end{eqnarray}
Again the quantities are restricted in order to get the instability
range as under
\begin{eqnarray}\nonumber&&
\left(\frac{ar}{r-m_0}\right)'+3\left(\frac{br}{r+m_0}\right)'
+\left(\frac{c}{r}\right)'<0, \quad
\left(\frac{ar}{r-m_0}\right)^\theta
+4\left(\frac{br}{r+m_0}\right)^\theta<0\\\nonumber&&
\left(\frac{br}{r+m_0}\right)'>0, \quad m_0<\frac{r}{2},
\end{eqnarray}
The self gravitating axially symmetric sources remains unstable in
pN-regime until the inequality $(4.46)$ remains valid which
analytically defines the instability range. The results for some
specific form of metric coefficients and the usual matter can be
deduced by the induction of corresponding values in dynamical
equations that further bring variations in adiabatic index
accordingly.

\section{Summary and Discussion}

The significance of dynamical analysis in modified gravity theories
urge us to explore the dynamical instability for axial symmetry of
self-gravitating systems in $f(R)$ framework. The motivation for the
study of axial symmetry came from the fact that the observational
gravitating systems may deviate from the most studied spherical
symmetry. Obviously it is not the basic characteristic of
gravitating sources, but such situation may occur incidently.
Herein, we deal with the restricted class of axially symmetric
sources, i.e., ignoring meridional motions and motions around
symmetry axis. Consequently, vorticity of sources with respect to
the system vanishes for observer at rest (vorticity-free case of
axial symmetry).

The $f(R)$ model, we have taken is the extension of extensively
studied Starobinsky model \cite{39a}, i.e. inclusion of $n^{th}$
order term of curvature in $f(R)=R+\alpha R^2$, provided that
$n\geq3$ \cite{40}. The extended Starobinsky model describe the
supersymmetric supergravity model constructed to include more
general analysis of the higher order curvature contributions. The
$f(R)$ form we have chosen is viable, satisfying both viability
criterion, i.e., positivity of first and second order derivatives.

The field equations for restricted axially symmetric sources with
three independent metric functions are formulated in $f(R)$ gravity.
The modified field equations are utilized to develop the dynamical
equation for the anisotropic fluid by consideration of contacted
Bianchi identities (conservation equations). The modified dynamical
equations are highly complicated non-linear equations whose general
solution has not been ascertained yet, that is why we have used
perturbation scheme to study the dynamical system under the
influence of extended Starobinsky model. Eulerian frame has been
considered for the dynamical analysis, initially all the physical
quantities are assumed to be in hydrostatic equilibrium.
Implementation of linear perturbation on dynamical equations assists
in the formation of collapse equation.

The collapse equation and substitution of the expressions for
perturbed energy density, pressure anisotropy leads to the second
order ordinary differential equations, which is used in estimation
of adiabatic index in terms of the material and dark source
components. The instability range for the N and pN regime has been
established in Eqs. $(4.45)$ and $(4.46)$, respectively. For each
approximation the physical quantities are constrained in order to
satisfy the stellar stable configuration. It is found that $R^n$
extension of Starobinsky model describes the more practical
substitute for higher order curvature corrections. The results are
analytic and so more generic, stability range for some particular
scenarios can be analyzed in depth by considering numerical
approach. The limiting case $\alpha\rightarrow0, \beta\rightarrow0$
defines the correction to GR solutions. The assumption
$\beta\rightarrow0$ corresponds to the dynamical analysis of
Starobinsky model, in accordance with \cite{39}.

\section*{Appendix}

The dark Source components of the perturbed dynamical equation are
given in following set of equations
\begin{eqnarray}\setcounter{equation}{1}\nonumber&&
Z_{1p}=\frac{e(1-\beta n(2-n)R_0^{n-1})}{2}-A_0^2\left\{\frac{1}{A_0^2B_0^2I^2}\left(
J'(1-\frac{b}{B_0})-\frac{A_0'}{A_0}J\right)\right\}_{,1}\\\nonumber&&
+\frac{LJ}{2I}-\frac{A_0^2}{r^2}\left\{\frac{1}{A_0^2B_0^2I^2}\left(
J^\theta(1-\frac{b}{B_0})-\frac{A_0^\theta}{A_0}J\right)\right\}_{,2}+\frac{1}{B_0^2}
\left[-\left(\frac{2a}
{A_0}+\frac{2b}{B_0}+\right.\right.\\\nonumber&&\left.\left.
\frac{J}{I}\right)\left\{
2\alpha\left((R_0R_0')'+\frac{(R_0R_0^\theta)^\theta}{r^2}\right)+\beta n(n-1)
\left((R_0^{n-2}R_0')'
+\frac{(R_0^{n-2}R_0^\theta)^\theta}{r^2}\right)\right\}\right.\\\nonumber&&\left.
+I'\left\{\left(\frac{c}{C_0}\right)'-2\left(\frac{b}{B_0}\right)'
-\frac{b}{B_0}\left(\frac{2A_0'}{A_0}+\frac{2B_0'}{B_0}-\frac{3}{r}\right)
-\frac{c}{C_0}\left(\frac{A_0'}{A_0}-\frac{C_0'}{C_0}-\frac{1}{r}\right)\right.\right.\\\nonumber&&\left.\left.+
\frac{J}{I}\left(\frac{C_0'}{C_0}+\frac{2B_0'}{B_0}-\frac{3}{r}\right)\right\}
+(e'J)'+\frac{(e^\theta J)^\theta}{r^2}
+\frac{I^\theta}{r^2}\left\{\left(\frac{c}{C_0}\right)^\theta
-2\left(\frac{b}{B_0}\right)^\theta-
\right.\right.\\\nonumber&&\left.\left.
\frac{b}{B_0}\left(\frac{2A_0^\theta}{A_0}+\frac{2B_0^\theta}{B_0}\right)
-\frac{c}{C_0}\left(\frac{A_0^\theta}{A_0}-\frac{C_0^\theta}{C_0}\right)+
\frac{J}{I}\left(\frac{C_0^\theta}{C_0}+\frac{2B_0^\theta}{B_0}\right)\right\}+J'\left(\frac{C_0'}{C_0}
\right.\right.\\\nonumber&&\left.\left.
-\frac{2B_0'}{B_0}+\frac{1}{r}\right)+\frac{J^\theta}{r^2}\left(\frac{C_0^\theta}{C_0}
-\frac{2B_0^\theta}{B_0}\right)+\left(\frac{2a}{A_0}+\frac{b}{B_0}\right)\left(I''
+\frac{I^{\theta\theta}}{r^2}\right)
-\left(\frac{3A_0'}{A_0}\right.\right.\\\nonumber&&\left.\left.
+\frac{C_0'}{C_0}
+\frac{2B_0'}{B_0}+\frac{1}{r}\right)\left(\frac{J'}{I}(1-\frac{b}{B_0})
-\frac{A_0'}{A_0}\frac{J}{I}\right)
+\left(\frac{3A_0^\theta}{A_0}+\frac{C_0^\theta}{C_0}
+\frac{2B_0^\theta}{B_0}\right)\left(\frac{A_0^\theta}{A_0}
\frac{J}{I}\right.\right.\\\label{a1}&&\left.\left.-\frac{J^\theta}{I}(1-\frac{b}{B_0})\right)\right],
\end{eqnarray}
\begin{eqnarray}\nonumber&&
Z_{2p}=\left[\left[\frac{1}{IB_0^2}\left\{\frac{\ddot{D}}{DA^2_0}-\frac{1}{B_0^2}
\left\{\frac{NB_0^2}{2}+I'\left(\left(\frac{a}{A_0}\right)'-\left(\frac{b}{B_0}\right)'
+\left(\frac{c}{C_0}\right)'\right)
\right.\right.\right.\right.\\\nonumber&&\left.\left.\left.\left.
+\left(J'-\frac{2b}{B_0}I'\right)\left(\frac{A_0'}{A_0}+\frac{C_0'}{C_0}
-\frac{B_0'}{B_0}-\frac{1}{r}\right)+\frac{1}{r^2}\left(J^{\theta\theta}+\left(J^\theta
-\frac{2b}{B_0}I^\theta\right)
\left(\frac{A_0^\theta}{A_0}
\right.\right.\right.\right.\right.\right.\\\nonumber&&\left.\left.\left.\left.\left.\left.-
\frac{3B_0^\theta}{B_0}
+\frac{C_0^\theta}{C_0}\right)+
\frac{2b}{B_0}I^{\theta\theta}+I^\theta
\left(\left(\frac{a}{A_0}\right)^\theta+\left(\frac{c}{C_0}\right)^\theta-3\left(\frac{b}{B_0}\right)^\theta
\right)\right)\right\}\right\}\right]_{,1}\right.\\\nonumber&&\left.+
\left[\frac{1}{r^2IB_0^4}\left\{J'^{\theta}+\left(\frac{b}{B_0}\right)^\theta I'+
J^{\theta}\left(\frac{B_0'}{B_0}+\frac{1}{r}\right)
-\left(\frac{b}{B_0}\right)'I^\theta\right\}\right]_{,2}\right]IB_0^4
\\\nonumber&&-N\frac{B_0'}{B_0}+\frac{A_0'}{A_0}\left[J''+\frac{J^{\theta\theta}}{r^2}
-\frac{2b}{B_0}\left(I''+\frac{I^{\theta\theta}}{r^2}\right)+\left(J'
-\frac{2b}{B_0}I'\right)\left(\frac{C_0'}{C_0}
-\frac{2B_0'}{B_0}\right.\right.\\\nonumber&&\left.\left.
+\frac{1}{r}\right)+I'\left(\left(\frac{c}{C_0}\right)'-\left(\frac{b}{B_0}\right)'
\right)+\frac{1}{r^2}\left\{I^\theta
\left(\left(\frac{c}{C_0}\right)^\theta-2\left(\frac{b}{B_0}\right)^\theta
\right)+\left(J^\theta\right.\right.\right.\\\nonumber&&\left.\left.\left.
-\frac{2b}{B_0}I^\theta\right)\left(\frac{C_0^\theta}{C_0}-\frac{2B_0^\theta}{B_0}\right)\right\}\right]+
\left(\frac{(aA_0)'}{A_0^2}-\frac{2b}{B_0}\frac{A_0'}{A_0}\right)\left(\frac{LB_0^2}{2}+I''+
I'\left(\frac{C_0'}{C_0}\right.\right.\\\nonumber&&\left.\left.
-\frac{2B_0'}{B_0}+\frac{1}{r}\right)+\frac{I^{\theta\theta}}{r^2}
+\frac{I^{\theta}}{r^2}\left(\frac{C_0^\theta}{C_0}-\frac{2B_0^\theta}{B_0}\right)\right)-
\left\{\frac{LB_0^2}{2}+I'\left(\frac{A_0'}{A_0}+\frac{C_0'}{C_0}
-\frac{B_0'}{B_0}\right.\right.\\\nonumber&&\left.\left.-\frac{1}{r}\right)+\frac{I^{\theta\theta}}{r^2}
+\frac{I^{\theta}}{r^2}\left(\frac{A_0^\theta}{A_0}+\frac{C_0^\theta}{C_0}-\frac{3B_0^\theta}{B_0}\right)\right\}
\left(\left(\frac{a}{A_0}\right)'+3\left(\frac{b}{B_0}\right)'
+\left(\frac{c}{C_0}\right)'\right)\\\nonumber&&-\left(\frac{A_0'}{A_0}+\frac{C_0'}{C_0}
+\frac{3B_0'}{B_0}+\frac{1}{r}\right)\left\{I'\left(\left(\frac{a}{A_0}\right)'-\left(\frac{b}{B_0}\right)'
+\left(\frac{c}{C_0}\right)'\right)
+\left(\frac{A_0'}{A_0}
-\frac{B_0'}{B_0}\right.\right.\\\nonumber&&\left.\left.
+\frac{C_0'}{C_0}-\frac{1}{r}\right)\left(J'-\frac{2b}{B_0}I'\right)+\frac{1}{r^2}\left(J^{\theta\theta}+\left(J^\theta
-\frac{2b}{B_0}I^\theta\right)
\left(\frac{A_0^\theta}{A_0}-
\frac{3B_0^\theta}{B_0}
+\frac{C_0^\theta}{C_0}\right)\right.\right.\\\nonumber&&\left.\left.+
\frac{2b}{B_0}I^{\theta\theta}+I^\theta
\left(\left(\frac{a}{A_0}\right)^\theta+\left(\frac{c}{C_0}\right)^\theta
-3\left(\frac{b}{B_0}\right)^\theta
\right)\right)\right\}
-\left[\left(\left(\frac{a}{A_0}\right)^\theta
+\left(\frac{c}{C_0}\right)^\theta
\right.\right.\\\nonumber&&\left.\left.+4\left(\frac{b}{B_0}\right)^\theta
\right)\left(I'^\theta +\frac{B_0^\theta}{B_0}I'+I^\theta\left(\frac{B_0'}{B_0}
+\frac{1}{r}\right)\right)-\left(\frac{A_0^\theta}{A_0}+
\frac{4B_0^\theta}{B_0}
+\frac{C_0^\theta}{C_0}\right)\left(\frac{B_0^\theta}{B_0}J'
\right.\right.\\\nonumber&&\left.\left.
-J'^\theta- I'\left(\frac{b}{B_0}\right)^\theta -J^\theta\left(\frac{B_0'}{B_0}
+\frac{1}{r}\right)+I^\theta\left(\frac{b}{B_0}\right)'\right)\right]\frac{1}{r^2}-
\left(\frac{B_0'}{B_0}+\frac{1}{r}\right)\left[\frac{B_0^2}{A_0^2}\frac{\ddot{D}}{D}J
\right.\\\nonumber&&\left.-J''+\frac{2b}{B_0}I''-I'\left(\left(\frac{a}{A_0}\right)'-\left(\frac{b}{B_0}\right)'
+\left(\frac{c}{C_0}\right)'\right)
-\left(J'-\frac{2b}{B_0}I'\right)\left(\frac{A_0'}{A_0}+\frac{C_0'}{C_0}\right.\right.
\end{eqnarray}
\begin{eqnarray}
\nonumber&&\left.\left.-\frac{B_0'}{B_0}\right)+\frac{1}{r^2}\left(I^\theta
\left(\left(\frac{a}{A_0}\right)^\theta+\left(\frac{c}{C_0}\right)^\theta-\left(\frac{b}{B_0}\right)^\theta
\right)-\left(J^\theta -\frac{2b}{B_0}I^\theta\right)
\left(\frac{A_0^\theta}{A_0}-\frac{B_0^\theta}{B_0}
\right.\right.\right.\\\nonumber&&\left.\left.\left.+\frac{C_0^\theta}{C_0}\right)\right)\right]
+\left(\frac{b}{B_0}\right)'\left[\frac{LB_0^2}{2}+I'
\left(\frac{A_0'}{A_0}+\frac{C_0'}{C_0}-\frac{B_0'}{B_0}\right)-\frac{I^\theta}{r^2}
\left(\frac{A_0^\theta}{A_0}-\frac{B_0^\theta}{B_0}
+\frac{C_0^\theta}{C_0}\right)\right.\\\nonumber&&\left.-I''\right]+
\frac{C_0'}{C_0}\left[J''-\frac{2b}{B_0}I''+I'\left(\left(\frac{a}{A_0}\right)'-\left(\frac{2b}{B_0}\right)'\right)
+\left(J'-\frac{2b}{B_0}I'\right)\left(\frac{A_0'}{A_0}-\frac{B_0'}{B_0}\right.\right.\\\nonumber&&\left.\left.
+\frac{1}{r}\right) +\frac{1}{r^2}\left\{J^{\theta\theta}+
\left(J^\theta -\frac{2b}{B_0}I^\theta\right)
\left(\frac{A_0^\theta}{A_0}-\frac{2B_0^\theta}{B_0}\right)+I^\theta
\left(\left(\frac{a}{A_0}\right)^\theta-2\left(\frac{b}{B_0}\right)^\theta
\right)\right.\right.\\\nonumber&&\left.\left.
-\frac{2b}{B_0}I^{\theta\theta}\right\}\right]+\left(\frac{(cC_0)'}{C_0^2}-\frac{2b}{B_0}\frac{C_0'}{C_0}\right)
\left[I'
\left(\frac{A_0'}{A_0}-\frac{2B_0'}{B_0}+\frac{1}{r}\right)-\frac{I^\theta}{r^2}
\left(\frac{A_0^\theta}{A_0}-\frac{2B_0^\theta}{B_0}\right)
\right.\\\label{a2}&&\left.+\frac{LB_0^2}{2}+I''+\frac{I^{\theta\theta}}{r^2}\right]-
\frac{\ddot{D}B_0^2}{DA^2_0I}\left(J'-\frac{A_0'}{A_0}J-\frac{b}{B_0}I'\right),
\\\nonumber&&
Z_{3p}=Ir^2B_0^4\left[\left[\frac{1}{r^2IB_0^4}\left\{J'^{\theta}+\left(\frac{b}{B_0}\right)^\theta
I'+ J^{\theta}\left(\frac{B_0'}{B_0}+\frac{1}{r}\right)
-\left(\frac{b}{B_0}\right)'I^\theta\right\}\right]_{,1}\right.\\\nonumber&&\left.+
\frac{\ddot{D}B_0^2}{DA^2_0I}\left(\frac{A_0^\theta}{A_0}J+\frac{b}{B_0}I^\theta-J^\theta\right)
+\left[\frac{1}{Ir^2B_0^4}\left\{\frac{\ddot{D}B_0^2}{DA_0^2}J
-J''+\frac{2b}{B_0}I''+\left(\frac{2b}{B_0}I'\right.\right.\right.\right.
\\\nonumber&&\left.\left.\left.\left.-J'\right)\left(\frac{A_0'}{A_0}+\frac{C_0'}{C_0}
-\frac{B_0'}{B_0}\right)-I'\left(\left(\frac{a}{A_0}\right)'-\left(\frac{b}{B_0}\right)'
+\left(\frac{c}{C_0}\right)'\right)
+\frac{1}{r^2}\left(\left(\frac{2b}{B_0}I^\theta
\right.\right.\right.\right.\right.
\\\nonumber&&\left.\left.\left.\left.\left.-J^\theta
\right)
\left(\frac{A_0^\theta}{A_0}-\frac{B_0^\theta}{B_0}+\frac{C_0^\theta}{C_0}\right)-I^\theta
\left(\left(\frac{a}{A_0}\right)^\theta+\left(\frac{c}{C_0}\right)^\theta-\left(\frac{b}{B_0}\right)^\theta
\right)\right)\right\}\right]_{,2}\right]
\\\nonumber&&-N\frac{B_0^\theta}{B_0}+\frac{A_0^\theta}{A_0}\left[J''+\frac{J^{\theta\theta}}{r^2}
-\frac{2b}{B_0}\left(I''+\frac{I^{\theta\theta}}{r^2}\right)+\left(J'
-\frac{2b}{B_0}I'\right)\left(\frac{C_0'}{C_0}
-\frac{2B_0'}{B_0}\right.\right.\\\nonumber&&\left.\left.
+\frac{1}{r}\right)+I'\left(\left(\frac{c}{C_0}\right)'-\left(\frac{b}{B_0}\right)'
\right)+\frac{1}{r^2}\left\{I^\theta
\left(\left(\frac{c}{C_0}\right)^\theta-2\left(\frac{b}{B_0}\right)^\theta
\right)+\left(J^\theta\right.\right.\right.\\\nonumber&&\left.\left.\left.
-\frac{2b}{B_0}I^\theta\right)\left(\frac{C_0^\theta}{C_0}-\frac{2B_0^\theta}{B_0}\right)\right\}\right]
+\left(\frac{(aA_0)^\theta}{A_0^2}-\frac{2b}{B_0}\frac{A_0^\theta}{A_0}\right)\left(\frac{LB_0^2}{2}+I''+
I'\left(\frac{C_0'}{C_0}\right.\right.
\end{eqnarray}
\begin{eqnarray}\nonumber&&\left.\left.
-\frac{2B_0'}{B_0}+\frac{1}{r}\right)+\frac{I^{\theta\theta}}{r^2}
+\frac{I^{\theta}}{r^2}\left(\frac{C_0^\theta}{C_0}-\frac{2B_0^\theta}{B_0}\right)\right)
-\left(\frac{b}{B_0}\right)^\theta\left\{\frac{LB_0^2}{2}+I'\left(\frac{A_0'}{A_0}+\frac{C_0'}{C_0}
\right.\right.\\\nonumber&&\left.\left.-\frac{B_0'}{B_0}+\frac{1}{r}\right)+\frac{I^{\theta\theta}}{r^2}
+\frac{I^{\theta}}{r^2}\left(\frac{A_0^\theta}{A_0}+\frac{C_0^\theta}{C_0}-\frac{3B_0^\theta}{B_0}\right)\right\}
-\frac{B_0^\theta}{B_0}\left\{I'\left(\left(\frac{a}{A_0}\right)'-\left(\frac{b}{B_0}\right)'
\right.\right.\\\nonumber&&\left.\left.
+\left(\frac{c}{C_0}\right)'\right) +\left(\frac{A_0'}{A_0}
-\frac{B_0'}{B_0}
+\frac{C_0'}{C_0}-\frac{1}{r}\right)\left(J'-\frac{2b}{B_0}I'\right)+\frac{1}{r^2}\left(J^{\theta\theta}
-\left(\frac{2b}{B_0}I^\theta-\right.\right.\right.\\\nonumber&&\left.\left.\left.J^\theta\right)
\left(\frac{A_0^\theta}{A_0}- \frac{3B_0^\theta}{B_0}
+\frac{C_0^\theta}{C_0}\right)+
\frac{2b}{B_0}I^{\theta\theta}+I^\theta
\left(\left(\frac{a}{A_0}\right)^\theta+\left(\frac{c}{C_0}\right)^\theta
-3\left(\frac{b}{B_0}\right)^\theta \right)\right)\right\}
\\\nonumber&&
-\frac{1}{r^2}\left[\left(\left(\frac{a}{A_0}\right)'
+\left(\frac{c}{C_0}\right)'+\left(\frac{b}{B_0}\right)'
\right)\left(I'^\theta +\frac{B_0^\theta}{B_0}I'+I^\theta\left(\frac{B_0'}{B_0}
+\frac{1}{r}\right)\right)-\left(\frac{A_0'}{A_0}\right.\right.\\\nonumber&&\left.\left.+
\frac{4B_0'}{B_0}
+\frac{C_0'}{C_0}\right)\left(\frac{B_0^\theta}{B_0}J'
-J'^\theta- I'\left(\frac{b}{B_0}\right)^\theta -J^\theta\left(\frac{B_0'}{B_0}
+\frac{1}{r}\right)+I^\theta\left(\frac{b}{B_0}\right)'\right)\right]
\\\nonumber&&+\left(\frac{A_0^\theta}{A_0}+
\frac{3B_0^\theta}{B_0}
+\frac{C_0^\theta}{C_0}\right)\left[\frac{B_0^2}{A_0^2}\frac{\ddot{D}}{D}J+
\frac{2b}{B_0}I''-I'\left(\left(\frac{a}{A_0}\right)'-\left(\frac{b}{B_0}\right)'
+\left(\frac{c}{C_0}\right)'
\right)\right.\\\nonumber&&\left.-J''
-\left(J'-\frac{2b}{B_0}I'\right)\left(\frac{A_0'}{A_0}+\frac{C_0'}{C_0}
-\frac{B_0'}{B_0}\right)-\frac{1}{r^2}\left(\left(J^\theta
-\frac{2b}{B_0}I^\theta\right)
\left(\frac{A_0^\theta}{A_0}-\frac{B_0^\theta}{B_0}
\right.\right.\right.
\\
\nonumber&&\left.\left.\left.+\frac{C_0^\theta}{C_0}\right)-I^\theta
\left(\left(\frac{a}{A_0}\right)^\theta+\left(\frac{c}{C_0}\right)^\theta-\left(\frac{b}{B_0}\right)^\theta
\right)\right)\right]+
\left(\left(\frac{a}{A_0}\right)^\theta+\left(\frac{c}{C_0}\right)^\theta
\right.\\\nonumber&&\left.+3\left(\frac{b}{B_0}\right)^\theta
\right)\left[\frac{LB_0^2}{2}+I'
\left(\frac{A_0'}{A_0}+\frac{C_0'}{C_0}-\frac{B_0'}{B_0}\right)-\frac{I^\theta}{r^2}
\left(\frac{A_0^\theta}{A_0}-\frac{B_0^\theta}{B_0}
+\frac{C_0^\theta}{C_0}\right)-I''\right]
\\\nonumber&&-\frac{C_0^\theta}{C_0}\left[J''-\frac{2b}{B_0}I''
+I'\left(\left(\frac{a}{A_0}\right)'-\left(\frac{2b}{B_0}\right)'\right)
+\left(J'-\frac{2b}{B_0}I'\right)\left(\frac{A_0'}{A_0}-\frac{B_0'}{B_0}\right.\right.\\\nonumber&&\left.\left.
+\frac{1}{r}\right)+\frac{1}{r^2}\left\{\left(J^\theta-\frac{2b}{B_0}I^\theta\right)
\left(\frac{A_0^\theta}{A_0}-\frac{2B_0^\theta}{B_0}\right)+I^\theta
\left(\left(\frac{a}{A_0}\right)^\theta-2\left(\frac{b}{B_0}\right)^\theta
\right)\right.\right.\\\nonumber&&\left.\left.
+J^{\theta\theta}-\frac{2b}{B_0}I^{\theta\theta}\right\}\right]+\left(\frac{(cC_0)^\theta}{C_0^2}
-\frac{2b}{B_0}\frac{C_0^\theta}{C_0}\right)
\left[\frac{LB_0^2}{2}+I'\left(\frac{A_0'}{A_0}
-\frac{2B_0'}{B_0}+\frac{1}{r}\right)
\right.\\\label{a3}&&\left.+I''+\frac{I^{\theta\theta}}{r^2}-\frac{I^\theta}{r^2}r
\left(\frac{A_0^\theta}{A_0}-\frac{2B_0^\theta}{B_0}\right)\right],
\end{eqnarray}
where, $N=e(1-\beta n(n-1))R_0^{n-1})$
\begin{eqnarray}\nonumber &&
Z_4=\frac{A_0^2}{2}\left(\frac{B_0C_0}{bC_0-cB_0}\right)\left[\frac{2}{B_0^2}
\left\{\frac{A_0'C_0'}{A_0C_0}\left(\frac{a'}{A'_0}-\frac{a}{A_0}+\frac{c'}{C'_0}
-\frac{c}{C_0}\right)+\frac{A_0''}{A_0}\left(\frac{a''}{A''_0}
\right.\right.\right.\\\nonumber &&\left.\left.\left.-\frac{a}{A_0}\right)
+\frac{B_0''}{B_0}\left(\frac{b''}{B''_0}-\frac{b}{B_0}\right)+\frac{C_0''}{C_0}\left(\frac{c''}{C''_0}
-\frac{c}{C_0}\right)-
\frac{1}{r}\left(\frac{a}{A_0}-\frac{b}{B_0}-\frac{c}{C_0}\right)'
\right.\right.\\\nonumber &&\left.\left.-\frac{2B_0'}{B_0}\left(\frac{b}{B_0}\right)'+\frac{2}{r^2}\left\{
\frac{2B_0^\theta}{B_0}\left(\frac{b}{B_0}\right)^\theta+
\frac{A_0^{\theta\theta}}{A_0}\left(\frac{a^{\theta\theta}}{A^{\theta\theta}_0}-\frac{a}{A_0}\right)
+\frac{B_0^{\theta\theta}}{B_0}\left(\frac{b^{\theta\theta}}{B^{\theta\theta}_0}
-\frac{b}{B_0}\right)\right.\right.\right.\\\label{a4} &&\left.\left.\left.+\frac{C_0^{\theta\theta}}{C_0}\left(\frac{c^{\theta\theta}}{C^{\theta\theta}_0}
-\frac{c}{C_0}\right)+\frac{A_0^\theta C_0^\theta}{A_0C_0}\left(\frac{a^\theta}{A^\theta_0}-\frac{a}{A_0}+\frac{c^\theta}{C^\theta_0}
-\frac{c}{C_0}\right)\right\}\right\}-e-\frac{2bR_0}{B_0}\right].
\end{eqnarray}

\end{document}